\documentclass[conference]{IEEEtran}
\IEEEoverridecommandlockouts
\usepackage{cite}
\usepackage{amsmath,amssymb,amsfonts}
\usepackage{algorithmic}
\usepackage{graphicx}
\usepackage{textcomp}
\usepackage{xcolor}
\usepackage{listings}

\newcommand{\eg}{{\it e.g.,}}

\begin{document}

\title{A Two-phase Recommendation Framework for Consistent Java Method Names \\
\thanks{}
}

\author{\IEEEauthorblockN{Weidong Wang}
\IEEEauthorblockA{\textit{Faculty of Information Technology} \\
\textit{Beijing University of Technology}\\
Beijing, China \\
wangweidong@bjut.edu.cn}
\and
\IEEEauthorblockN{Dian Li}
\IEEEauthorblockA{\textit{Faculty of Information Technology} \\
\textit{Beijing University of Technology}\\
Beijing, China \\
iamlidian@emails.bjut.edu.cn}
\and
\IEEEauthorblockN{Yujian Kang}
\IEEEauthorblockA{\textit{Faculty of Information Technology} \\
\textit{Beijing University of Technology}\\
Beijing, China \\
kangyujian@gmail.com}
}

\maketitle

\begin{abstract}
	
In software engineering (SE) tasks, the naming approach is so important that it attracts many scholars from all over the world to study how to improve the quality of method names. To accurately recommend method names, we employ a novel framework to handle this problem. In our expeirments, nearly 8 million Java methods are collected from open source organizations as our evaluation dataset. In the first-phase recommendation, we introduce a fast and simple classifier based on the fast text neural network for reccomending potential method category. In the second-phase recomendation, we employ both two Long Short Term Memory Networks to reccomend consitent method names from each classification. Evaluation results prove that the proposed approach significantly outperforms state-of-the-art approach. 

\end{abstract}


\section{Introduction}
\label{sec1}

In the software engineering field, it is hard to automatically generate high-quality method name including answering why an accurate method name occurring frequently requires synthesis of different kinds of knowledge and context. This paper is to learn the characteristics of method bodies by considering correlations of multiple method characteristics. Hence, we propose a novel framework to name methods. The contributions of this paper can be summarized as follows.

Firstly, to classify prefixes of method names, we introduce a novel feed-forward neural network model. Compared with traditional models, our customized model typically considers correlations among method body characteristics (\eg\ class name, return value type, and parameters) according to the characteristics of method.

Secondly, to suggest method names, we employ both two Long Short Term Memory Networks \cite{shi2015convolutional} in each classification. Unlike traditional recommendation approaches, the model is combined with multiple characteristics such as class name, method body, and interface, which can ensure the accuracy.

The rest of this paper is organized as follows. Section \ref{sec2} describes the details of the proposed framework. Section \ref{sec3} describes the experiment results. Section \ref{sec4} describes the discussion of related work. Section \ref{sec5} concludes this paper and outlines the future work.

\section{Two-phase Recommendation Framework}
\label{sec2}

This section introduces our framework. The framework of the proposed approach is shown in Figure 1. In the first phase, method names within the common prefixes from the open source organizations and databases are classified based on their method body characteristics. Based on the work, classification scheme associates each prefix to a particular name group (\eg\ set, get, is, and test). We further extend the feed-forward neural network as a classifier for the prefixes of method names according to the characteristics of method bodies. Compared with the previous work, the proposed approach can achieve more accurate classification results by considering the correlations among parameters' type, class name, and return type etc.

\begin{figure*}[t]
\centering
\includegraphics[width=15cm]{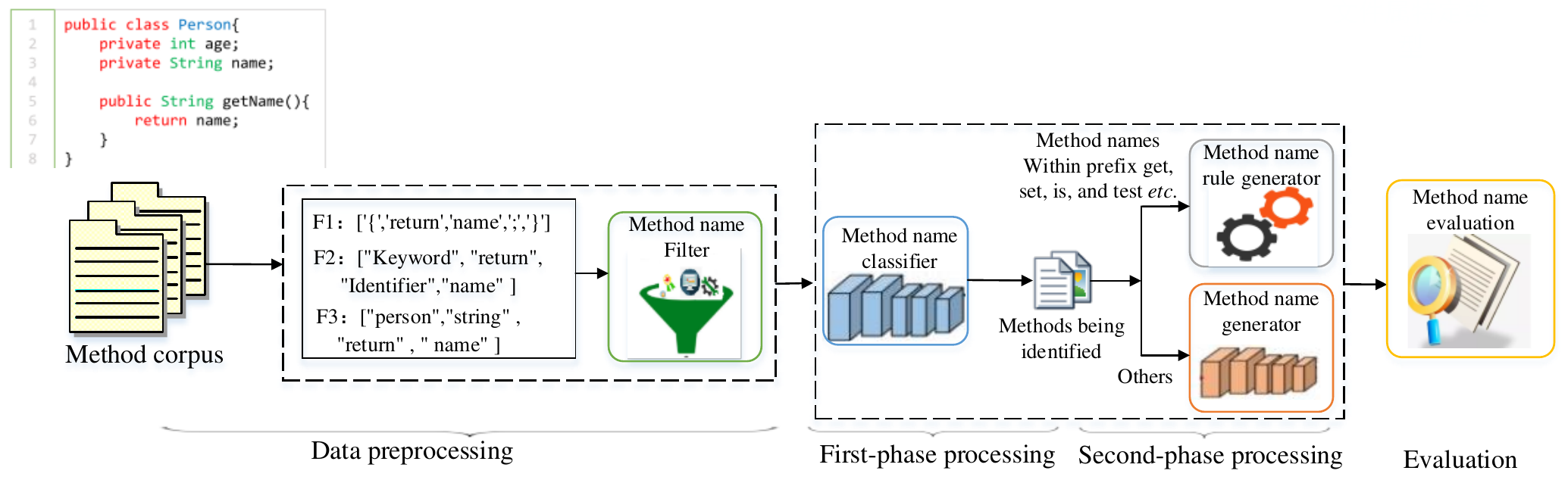}
\caption{The framework of efficient method name recommendation.}
\label{fig-1}
\end{figure*}

In the second phase, for each category, we employ LSTM network model \cite{g43,g45} to generate method names by considering the correlations among the characteristics of method  bodies. Compared with traditional naming methods based approaches, our approach explores the following capacities. 1) As the number of testing samples increases, our approach has high stable performance. Since it not only considers the correlations among the characteristics of method bodies but also depends on the external class name and context. 2) The two-phase neural networks can avoid noises from dataset such as large number of methods within the same prefix because the first-phase processing for classification can guarantee accurate ranking and the second-phase processing can benefit from the previous phase and obtain the primary characteristics of method bodies from different groups. More details will be described in the following sections.

\subsection{First-phase recommendation for Classification}
\label{subsec2-2}

The purpose of method classification is to classify method names within the prefixes (\eg\ set, get, is, test, others.) into different categories. To classify method names within such specific prefixes, we employ the \emph{Fasttext} \cite{g32} as the classifier in the first-phase processing because it speeds up training and testing processes while maintaining high precision.

\begin{figure}[t]
\centering
\includegraphics[width=8cm]{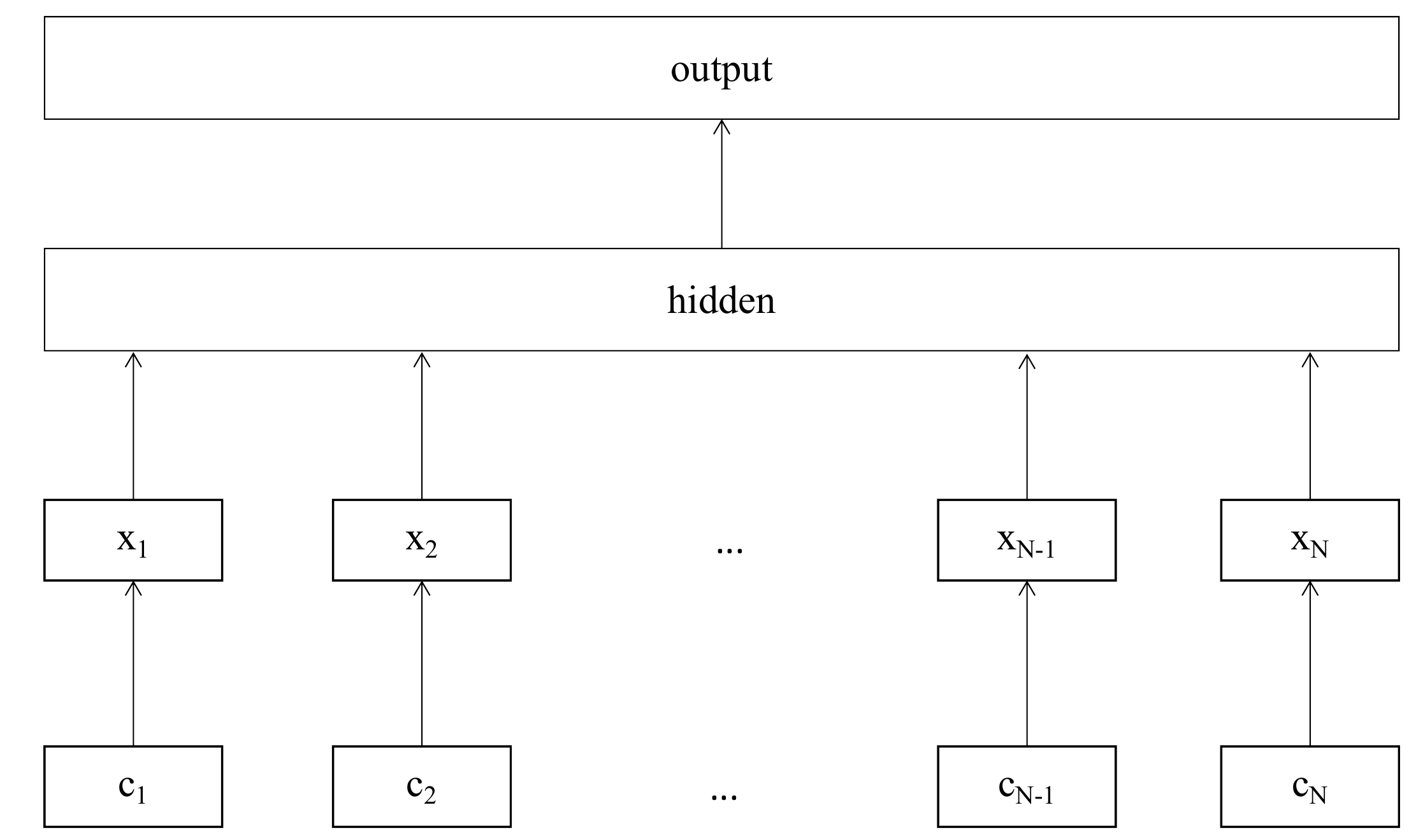}
\caption{Fasttext model architecture.}
\label{fig:1-1}
\end{figure}

As shown in Figure \ref{fig:1-1}, we introduce $ F_2$ format data $ c=\{c_1,c_2,c_3,...,c_n\}$ as the input of model. For the purpose of text vectorization, $c$ can be transformed into the vector representation $ x=\{x_1,x_2,x_3,...,x_n\}$. We use the Softmax function \cite{g35,g40} $f$ located in the hidden layer to calculate the probability distribution of the predefined categories. Specially, we train the model at the word level using bigram \cite{g36} value that is the concatenation of two consecutive tokens \cite{g34}. For a set of $N$ tokens, this leads to minimizing the negative log-likelihood as follows.

\begin{equation}
\footnotesize
-\frac{1}{N}\sum_{i=1}^{N}y_nlog(f(BAx_n)),
\label{formu033}
\end{equation}

\noindent where $ x_n $ is the normalized bag of features of the $n$-th token, $ y_n $ is the label, $ A $ and $ B $ are the weight matrixes. More details about Fasttext model can be found in literature \cite{g33}.

\subsection{Second-phase recommendation for consistent Method Names }
\label{subsec2-3}

The method classifiers divide Java methods into two categories. The first category consists of methods with special prefixes, while the second category includes methods without special prefixes. Here, we suppose that a prefix of method name including one of ``get'', ``set'', ``is'' or ``test'' is the first category. This is because the method names with these special prefixes account for up to 42\% of the total number of methods. For the first category, we use a heuristic rule for suggesting method names. For the second category, we introduce a neural network to recommend method names. The details are as follows:

\begin{figure}[b]
\centering
\includegraphics[width=9cm]{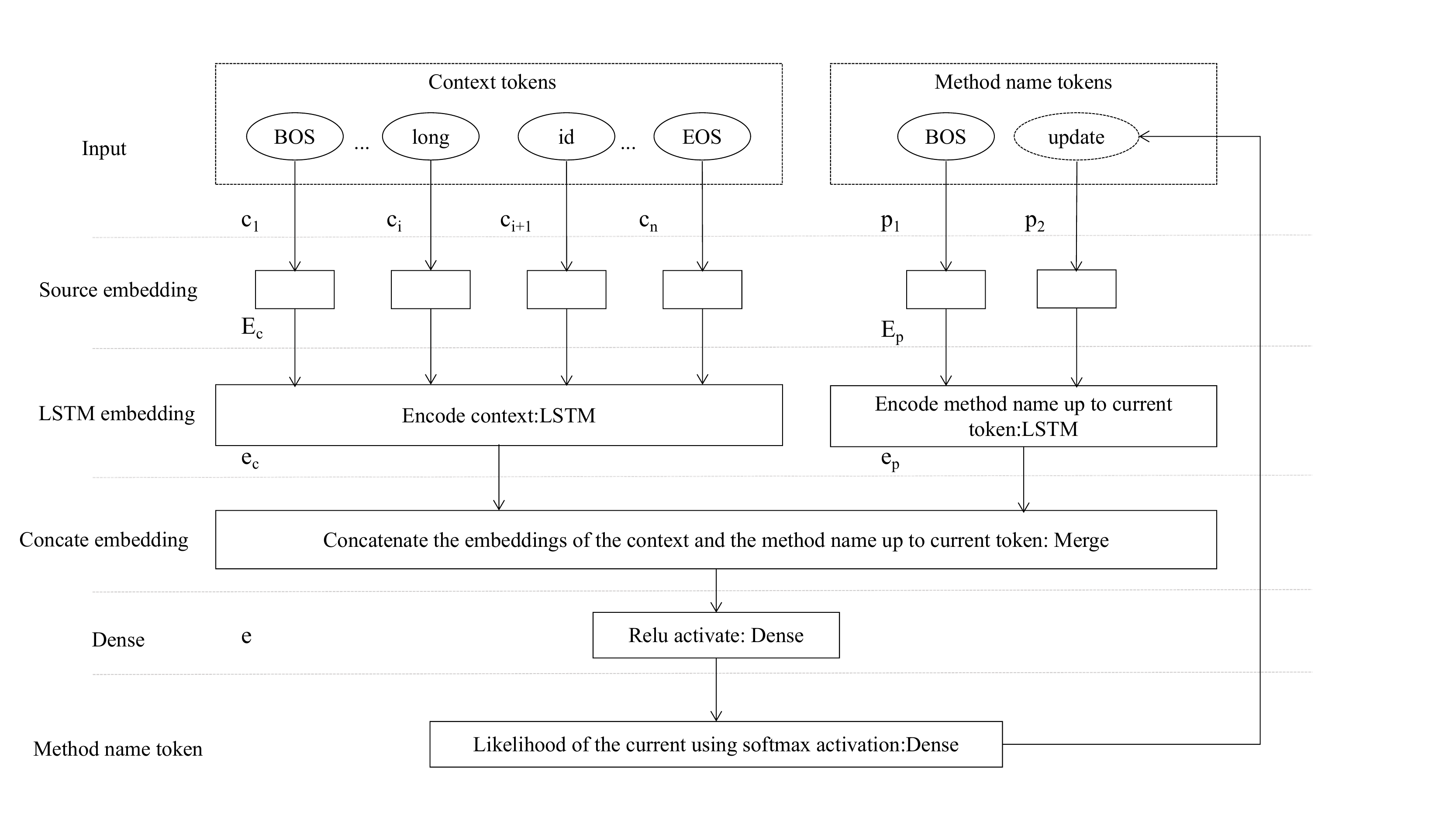}
\caption{The second-phase processing for generating method names.}
\label{fig:2-22}
\end{figure}

In the recommedation, the two Long Short Term Memory Networks (LSTM) \cite{g37} are adopted, both of which have the same architecture. One of them is responsible for handling $ E_c $ and the other is answerable for handling $ E_p $. The details of LSTM mathematics can be found in \cite{g39}. Then our two LSTMs are simply expressed as functions $ \Gamma_c $ and $ \Gamma_p $. These two functions are responsible for extracting the sentence embedding vector of context sequence and unfinished method name token sequence $ e_c $ and $ e_p $. $ \omega_c $ and $ \omega_p $ represent the parameters of two LSTMs separately.

\section{Experiments}
\label{sec3}

To evaluate the performance of such naming approaches, we conducted a series of experiments. The following parts mainly consist of the parameter settings of experiments, the choices of evaluation metrics, and the answers to the research questions from RQ1 to RQ2. Before describing the results and drawing conclusions, we first introduce the collected datasets and detailed parameter settings.

1) In order to study the advanced approaches to Java method name recommendation in this paper, we propose the following research questions (RQs).

2) To make the contrast experiment more objective and fair, we built 3 datasets for the purpose of experimental evaluation namely ``Liu-kui dataset", ``new-100 dataset", and ``new-400 dataset", respectively. The details of these datasets are in our previous work \cite{wang2022empirical} In the envaluation, we adopt 10-cross validation. 

\begin{table}[h]
\center
\begin{tabular}{|c|c|c|}
\hline
\multicolumn{3}{|c|}{[Dataset 1] Liu-kui Dataset} \\ \hline
Test          & Training          & Total            \\ \hline
211641        & 1904772        & 2,116,413        \\ \hline
\multicolumn{3}{|c|}{[Dataset 2] New-100 Dataset} \\ \hline
Test          & Training          & Total            \\ \hline
4710          & 42388          & 47098            \\ \hline
\multicolumn{3}{|c|}{[Dataset 3] New-400 Dataset} \\ \hline
Test          & Training         & Total            \\ \hline
481039        & 4329351        & 4810390          \\ \hline
\end{tabular}
\caption{The details of training and test samples in the three datasets.}
\label{tab-1}
\end{table}

\noindent \textbf {RQ1: Compared with the current advanced technologies to method name recommendation, what is the performance of the \emph{HeMa, code2vec, code2seq, and the framework proposed}?}

Compared with the current advanced technologies, the three popular approaches are empirically studied as follows. The first model is \emph{code2seq}, which compresses each AST path into a fixed length vector with LSTM, and uses the attention mechanism to select the relevant paths during decoding to generate each token in the method name. The second model is \emph{HeMa}, which employs heuristic rules to recommend method names. The third model is \emph{code2vec}, which extracts the AST structure of the method, then transforms the AST structure into the corresponding vector, finally generates the method name via the vector.

In order to compare the performance of the three popular approaches, we evaluated them on the Liu-kui dataset mentioned earlier. The aforementioned precision, recall, and F1-score are employed as evaluation metrics. The evaluation results are shown in table \ref{tab:10}.

\begin{table}[h]
\center
\begin{tabular}{|c|c|c|c|c|}
\hline
\textbf{} & \textbf{code2seq} & \textbf{code2vec} & \textbf{HeMa} & \textbf{The framework}  \\ \hline
\textbf{Precision} & 0.70     & 0.49     & 0.45 & 0.72 \\ \hline
\textbf{Recall}    & 0.63     & 0.34     & 0.38 & 0.64 \\ \hline
\textbf{F1-score} & 0.66     & 0.40     & 0.28 & 0.67 \\ \hline
\end{tabular}
\caption{The performance of different naming approaches.}
\label{tab:10}
\end{table}

As shown in Table \ref{tab:10}, the recall and precision of the framework exceed \emph{code2seq}, \emph{HeMa}, and \emph{code2vec} by 2\% to 27\% respectively.  First, for the recall, the sequence corresponding to the method name generated by the framework covers more correct tokens than the sequence corresponding to the method name generated by \emph{code2seq}. Second, because the framework has a higher precision, the proportion of correct tokens in the corresponding sequence of generated method names exceeds that of \emph{code2seq}.

\noindent \textbf{RQ2: Is our proposed approach effective for real developers?}

To answer RQ2, we investigate how the framework performs when it is really needed. To ensure that the results are as objective as possible, we also calculate the objective indicators corresponding to the method name generated by each method on different difficulty coefficient methods. As shown in table \ref{tab:15}, the first column is the developer's rating of the method naming difficulty, and the second to fourth columns are the precision, recall, and F1-score (F1) of the method name generated by each method on the method with corresponding difficulty coefficient.

\begin{table*}[t]
\center
\begin{tabular}{|c|c|c|c|c|c|c|c|c|c|c|c|c|}
\hline
 &
  \multicolumn{3}{c|}{\textbf{HeMa (\%)}} &
  \multicolumn{3}{c|}{\textbf{code2vec (\%)}} &
  \multicolumn{3}{c|}{\textbf{code2seq (\%)}} &
  \multicolumn{3}{c|}{\textbf{the framework (\%)}} \\ \hline
\textbf{Difficulty} &
  \multicolumn{1}{c|}{\textbf{recall}} &
  \multicolumn{1}{c|}{\textbf{precision}} &
  \multicolumn{1}{c|}{\textbf{F1}} &
  \multicolumn{1}{c|}{\textbf{recall}} &
  \multicolumn{1}{c|}{\textbf{precision}} &
  \multicolumn{1}{c|}{\textbf{F1}} &
  \multicolumn{1}{c|}{\textbf{recall}} &
  \multicolumn{1}{c|}{\textbf{precision}} &
  \multicolumn{1}{c|}{\textbf{F1}} &
  \multicolumn{1}{c|}{\textbf{recall}} &
  \multicolumn{1}{c|}{\textbf{precision}} &
  \multicolumn{1}{c|}{\textbf{F1}} \\ \hline
1(\textbf{Very Easy}) &
  95 &
  94 &
  94 &
  67 &
  69 &
  68 &
  93 &
  94 &
  93 &
  97 &
  95 &
  96 \\ \hline
2 (\textbf{Easy}) &
 38 &
 59 &
 46 &
 22 &
 28 &
 25 &
 69 &
 64 &
 66 &
 70 &
 65 &
 67 \\ \hline
3 (\textbf{Normal}) &
  50 &
  48 &
  49 &
  14 &
  15 &
  14 &
  48 &
  36 &
  41 &
  50 &
  40 &
  44 \\ \hline
4 (\textbf{Difficult}) &
  15 &
  16 &
  16 &
  17 &
  21 &
  19 &
  33 &
  24 &
  27 &
  27 &
  30 &
  28 \\ \hline
5 (\textbf{Extremely Difficult}) &
  7 &
  7 &
  7 &
  15 &
  16 &
  16 &
  55 &
  34 &
  42 &
  43 &
  45 &
  44 \\ \hline
\textbf{Final results} &
  67 &
  69 &
  68 &
  47 &
  47 &
  45 &
  76 &
  70 &
  73 &
  78 &
  73 &
  75 \\ \hline
\end{tabular}
\caption{The performance of different approaches in different difficulty levels of method naming.}
\label{tab:15}
\end{table*}

As shown in Table \ref{tab:15}, we have the following findings:

Above all, the heuristic approaches are better than those approaches based on machine learning in solving the problem of easy method name recommendation. Oppositely, the machine-learning based approaches perform better in naming difficult methods. By considering both advantages of them, the heuristic rules and machine-learning approaches are both employed by two-phase neural networks which recommends method names in different ways for different methods to obtain high efficiency.

\section{Discussion of Related Work}
\label{sec4}

Liu et al. \cite{g2} embedded method names and method bodies into numerical vectors using paragraph vector and \emph{word2vec} \emph{CNNs}, respectively. For a given method, they captured the similarity of method name vectors (\emph{NS} set) and method body vectors (\emph{BS} set). If $BS$ $\bigcap$ $NS$ $==$ $\emptyset$, the given method should be renamed, and a consistent method name would be suggested. 

Moreover, their approach helps developers to fix 66 inconsistent method names. Finally, they developed their method inconsistency detection tool and opened the data used in their study. JSNeat \cite{g8} introduced an information retrieval (IR)-based approach to search the names in minified code. JSNice \cite{g9} solved two kinds of problems in the context of JavaScript: predicting names of identifiers and predicting type annotations of variables.

Allamanis et al. \cite{g4} introduced an attentional neural network that employs convolution on the input tokens to solve the problem of extreme summarization of source code into short, descriptive function name-like summaries. Moreover, Allamanis's model \cite{g5} embedded the context for learning method bodies as tokens and put them with all the tokens of method names together for the same vector space. From the vector space, their model captured the semantically similar tokens to compose a new method name.

Code representations using neural networks for SE tasks have been proposed by several researchers \cite{g26}. The \emph{code2vec} \cite{g11,g13} utilizes the \emph{CNN} to construct code snippet embeddings for frequent paths on the ASTs. Deep learning similarity used recursive encoders on identifiers and ASTs. DeepSim \cite{g7} encoded code control flow and data flow into a semantic matrix and a neural network measures code functional similarity.

The other neural-network based classification approaches are for specific SE tasks including \emph{Fasttext}\cite{g14,g44}, \emph{TextCNN}\cite{g15}, \emph{Bert}\cite{g16}, \emph{TextRNN}\cite{g17}, \emph{RCNN}\cite{g18}. In addition, the recursive structure \cite{g41} is commonly found in the inputs of natural language sentences or code snippets. 

\section{Conclusions and Future Work}
\label{sec5}

Method names are critical for readability and maintainability of programs. In this paper, we propose a two-phase recommendation framework to efficiently generate method names according to the given method bodies. The experimental results prove that our proposed approach is accurate and efficient.

Future work is needed to investigate the professional developers' opinions on method naming standards.

\bibliographystyle{IEEEtran}
\bibliography{Reference}


\end{document}